\documentclass[aps,twocolumn,showpacs,showkeys]{revtex4}
\usepackage{bm,bbm}
\usepackage{graphicx}
\usepackage{amsmath}
\usepackage{amsfonts}
\usepackage{amssymb}
\usepackage[scannall]{psfrag}

\usepackage{latexsym}
\usepackage{color}

\newcommand{\eca}{\frac{\eta_2 \eta_3}{\eta_1}}
\newcommand{\ecb}{\frac{\eta_3 \eta_1}{\eta_2}}
\newcommand{\ecc}{\frac{\eta_1 \eta_2}{\eta_3}}

\begin{document}

\title{Unitary quantum gates, perfect entanglers
 and  unistochastic maps}

\author{Marcin Musz$^{1}$, Marek Ku{\'s}$^{1}$, and
     Karol \.Zyczkowski$^{1,2}$
\smallskip \\
$^1${\small Center for Theoretical Physics, Polish Academy of Sciences,
 Al. Lotnik\'ow 32/44, 02--668 Warszawa, Poland} \\
$^2${\small Institute of Physics, Jagiellonian University,
           ul. Reymonta 4, 30--059 Krak\'ow, Poland} \\
}

 \email{  marek.kus@cft.edu.pl, karol@cft.edu.pl}

% \date{January 29, 2013}

\begin{abstract}
Non-local properties of ensembles of quantum gates induced by the Haar
measure on the unitary group are investigated. We analyze the entropy of
entanglement of a unitary matrix $U$ equal to the Shannon entropy of the
vector of singular values of the reshuffled matrix. Averaging the entropy
over the Haar measure on $U(N^2)$ we find its asymptotic behaviour. For
two--qubit quantum gates we derive the induced probability distribution
of the interaction content and show that the relative volume of the set
of perfect entanglers reads $8/3 \pi \approx 0.85$. We establish explicit
conditions under which a given one-qubit bistochastic map is
unistochastic, so it can be obtained by partial trace over a one--qubit
environment initially prepared in the maximally mixed state.
\end{abstract}

 %Foundations of quantum mechanics, measurement theory
%\keywords{Quantum gates, non-local unitary operations}
\maketitle

%%====================================================================
\section{Introduction}
%%====================================================================

Unitary quantum gates form key ingredients of any quantum algorithm, so they
are widely used in the theory of quantum information \cite{NC00}. A unitary
gate acting on a bipartite system $A \otimes B$ is called {\sl local}, if the
unitary matrix has a form of the tensor product, $U=U_A \otimes U_B$. To
process quantum information between both subsystems one needs to use {\sl
non--local} gates, which are not of the product form.

Classification of unitary quantum gates is a subject of a considerable
interest \cite{KBG01,HVC02,ZVWS02,WSB03}. Quantification of the non-local
properties of unitary gates has been initiated by Zanardi and co-workers
\cite{ZZF00,Za01,WZ02}, while several other measures of non--locality were
introduced and analyzed in a seminal paper of Nielsen et al.
\cite{NDDGMOBHH02}. Since local unitary gates cannot produce quantum
entanglement, the non-local properties of a given gate $U$ may be
characterized by the average (or maximal) degree of entanglement of a
transformed separable state $|\psi'\rangle=U|\psi_{\rm sep}\rangle$. For
instance, the average linear entropy of a random product state transformed by
a bi-partite unitary gate leads to its {\sl entangling power}, introduced by
Zanardi et al. \cite{ZZF00}, and later investigated for various models in
\cite{SC03,WEP03,Sc04,BCPP05, CGSS05,LSW09,SM10}.
An alternative approach to the problem of non--locality based on the 
mimimal Frobenius distance of an analyzed global unitary matrix
to the closest local gate was recently discussed in \cite{Sa11},
where a relation to matrix product operator formalism
was established.

A given unitary gate $U$ is called a {\sl perfect entangler}, if there exists
a separable state transformed by $U$ into a maximally entangled state
\cite{Ma02}. Another class of maximally entangling unitary gates
was characterized in \cite{YGC10,Co11}.

Two unitary gates are called {\sl locally equivalent} if they coincide up to
local transformations. The  general problem of finding necessary and
sufficient conditions for local equivalence remains open. The full answer is
known in the simplest case of a two--qubit system, since a {\sl
canonical form} of a unitary gate of size four is established and any gate
can be uniquely described by a three--components vector called {\sl
information content} \cite{KBG01,KC01,DC02,DVC02}.

In this work we analyze properties of a 'typical' quantum gate. In other
words, we are going to average quantities characterizing each gate over a
unique, unitarily invariant, Haar measure on the space of unitary matrices.
An ensemble generated according to this measure is often referred to as {\sl
circular unitary ensemble} (CUE), the spectrum belongs to the unit circle and
the ensemble is invariant with respect to unitary transformation. An
exemplary algorithm of generating unitary matrices from this ensemble was
discussed in \cite{PZK98}. To study the set of quantum gates we found it
useful to define a {\sl special circular unitary ensemble} (SCUE) containing
special matrices with determinant equal to unity.

One of the main result of this paper consists in deriving the probability
distribution $P({\vec a})$ for the information content of a random two--qubit
gate, induced by the Haar measure on the unitary group. As an application of
this result we compute the relative volume of the set of perfect entanglers
with respect to this natural measure. Basing on numerical results performed
for unitary gates of larger dimensionalities we are in position to predict
asymptotic behavior of the average entanglement entropy of a random unitary
gate.

Furthermore, we analyze the class of {\sl unistochastic operations},
introduced in \cite{ZB04}, which can be described by a coupling with an
$M$--dimensional environment initially in the maximally mixed state,
\begin{equation}
\rho'=\Phi_U(\rho)={\rm Tr}_{\rm env}
 \bigl[ U \bigr(\rho \otimes \frac{\mathbbm 1}{N} \bigl) U^{\dagger} \bigr]
\label{unistoch}
\end{equation}
The partial trace is performed over the environment described in the Hilbert
space ${\cal H}_M$. If the dimension $N$  of the principal system and the
dimension $M$ of the ancillary system are equal, the map is called
unistochastic, while it is called  {\sl $k$-unistochastic} if $M=kN$. Thus a
unistochastic  map $\Psi_U$  is determined by a unitary matrix $U$  of size
$N^2$, while any $k$-unistochastic map is given by a matrix of size
$N^{k+1}$.

By construction any unistochastic  map is bistochastic, since the maximally
mixed state is preserved, $\Psi_U({\mathbbm 1}/N)={\rm Tr}_E ({\mathbbm
1_{N^2}})/N^2={\mathbbm 1_{N}}/N$. On the other hand, the converse is not
true, and in this paper we determine sufficient and necessary condition for a
one-qubit bistochastic map to be unistochastic.

The name of this class of maps is related to the classical case, in which
probability vectors are transformed by stochastic matrices. The matrix is
called bistochastic (or doubly stochastic), if it preserves the uniform
(maximally mixed) probability vector. A bistochastic matrix $B$ is called
{\sl unistochastic} if there exists an unitary $V$ of the same size  such
that $B_{ij}=|V_{ij}|^2$. For $N=2$ all bistochastic matrices are
unistochastic, but this is not the case already for $N=3$. For higher
dimensions the problem finding necessary and sufficient conditions for
unistochasticity remains open \cite{ZKSS03}.

In analogy to the classical case, any quantum map  determined as in
(\ref{unistoch}) by an orthogonal matrix will be called {\sl
orthostochastic}. In  \cite{ZB04} it was shown that for any unistochastic map
the spectrum of the corresponding dynamical matrix is given by the Schmidt
 coefficients of the unitary matrix $U$ treated as an element of the composite
Hilbert--Schmidt space. This implies that the entropy of such an operation
$S(\Psi_U)$ is equal to the entanglement entropy of the unitary matrix. In
other words, a link between non--local properties of a unitary gate acting on
a bipartite system and the decoherence induced by an associated unistochastic
map acting on a single system can be established.

It is appropriate to mention that the class of quantum maps for which the
system is coupled with the environment in the maximally mixed state was
already investigated in the literature. Such maps were discussed in the
context of quantum information processing \cite{KL98,PBLO03}, and, under the
name 'noisy maps', while studying reversible transformations from pure to
mixed states \cite{HHO03}. Moreover, Haagerup and Musat \cite{HM10}  analyzed
properties of a related class of {\sl factorizable} quantum maps introduced
in \cite{An06}. In fact $k$--unistochastic operations coincide with a subset
of these quantum maps called {\sl exactly factorizable}. In general this set
is not convex, and its convex hull defines a larger set of maps called {\sl
strongly factorizable}.

Note that a given unitary matrix $U$ of a composite dimension $d=N^2$ may
play very different roles in the theory of quantum information. Let us
specify here three most natural applications:

\smallskip
a) $U\in U(N^2)$ describes a quantum gate acting on a $N \times N$ bi-partite system, and
its operator Schmidt decomposition characterizes the non-local properties
\cite{KC01,HVC02,ZVWS02,NDDGMOBHH02},

\smallskip
b) $U\in U(N^2)$  determines by Eq. (\ref{unistoch})
   a unistochastic quantum operation, $\rho'=\Phi_U(\rho)$,
   acting on an $N$--level system   \cite{ZB04},
 %5 ={\rm Tr}_E [U (rho \otimes {\mathbbm I}/N) U^{\dagger}]$

\smallskip
c) $U\in U(N^2)$  defines a maximally entangled state of a composite,
$N^2\times N^2$ system \cite{We01,WGC03}, as
$|\psi\rangle = (U\otimes {\mathbbm I}) |\psi_+\rangle$
where $|\psi_+\rangle  = \frac{1}{N}\sum_{j=1}^{N^2} |j,j\rangle$.
\medskip

The paper is organized as follows. Unistochastic operations are analyzed in
section II. As any one--qubit unistochastic map is determined by a unitary
matrix of order four, we analyze in section III the set of all two--qubit
unitary gates. This allows us to characterize the set ${\cal U}_2$ of
one--qubit unistochastic maps, which forms a non-convex subset of the
tetrahedron of bistochastic maps spanned by three Pauli matrices and the
identity map. In section IV the ensemble of random two-qubit quantum gates is
described. It is based on circular unitary ensemble of special unitary
matrices of size four. The probability distribution of purity (nonlocality)
for this ensemble is computed. Furthermore, we derive the probability that a
generic gate belongs to the class of {\sl perfect entanglers}, so it can
transform a product state into a maximally entangled Bell-like  state.
Unitary gates acting on $N \times N$ systems are investigated in Section V.
For completeness, some basic properties of the operator Schmidt decomposition
and related algebra of reshuffling of a matrix are reviewed in Appendix A.

%%====================================================================
\section{Unistochastic maps}
\label{sec:unistomaps}
%%====================================================================

Any unitary matrix $U$ of size $N^2$ describes a unitary gate acting on the
bi--partite system. Alternatively it may be used to define a {\sl
unistochastic map}  \cite{ZB04} acting on a single system of size $N$
according to eq. (\ref{unistoch}).
In other words, the principal system is coupled to the ancilla of the same
size, $M=N$, prepared initially in the maximally mixed state. Unless the gate
$U$ is local so that $U=U_a \otimes U_b$, the partial trace leads to a
non-unitary evolution of the density matrix $\rho$.

Any such discrete map can be written in the {\sl Kraus form} \cite{Kr71}
\begin{equation}
  \rho' = \Phi(\rho) =  \sum_{i=1}^k A_i \rho  A_i^{\dagger} \ .
 \label{kraus}
\end{equation}
To preserve the trace, Tr$\rho'=$Tr$\rho=1$, the Kraus operators
need to satisfy the completeness relation
\begin{equation}
   \sum_{i=1}^k  A_i^{\dagger} A_i = {\mathbbm 1} \ .
 \label{complet}
\end{equation}
A trace preserving map $\Phi$ written in the form (\ref{kraus}) is called
{\sl stochastic}. It can be represented by the dynamical matrix
$D_{\Phi}=N(\Phi \otimes {\mathbbm I}) |\psi_+\rangle \langle \psi_+|$, also
called {\sl Choi matrix}.

To perform the partial trace over the environment in the definition
(\ref{unistoch}) let us apply the operator Schmidt decomposition of
$U$ recalled in (\ref{VSchmidt}). 
making use the notation introduced in Appendix the map  $\Phi$ can be
rewritten as
\begin{eqnarray}
\label{unitevol5}
&\rho' =  {\Phi}_U\rho =  {\rm Tr}_{\rm env} \Bigl[U(\rho\otimes
\frac{1}{N} {\mathbbm 1}_N)\, U^{\dagger} \Bigl]
%\quad\quad\quad \quad \quad
\nonumber
\\
 & {\displaystyle = {\rm Tr}_{\rm env} \Bigl[ \sum_{i=1}^{N^2}
\sum_{j=1}^{N^2} \sqrt{\Lambda_i\Lambda_j}
\bigl( B^{\prime}_i \rho B^{\prime \dagger}_j \bigr) \otimes
\bigl( \frac{1}{N} B^{\prime\prime}_i B^{\prime\prime\dagger}_j  \bigr) \Bigr]
} \nonumber
\\
 & = \frac{1}{N} \sum_{i=1}^{N^2} \Lambda_i B^{\prime}_i
\rho B^{\prime \dagger}_i \ .
\end{eqnarray}

The standard Kraus form is obtained by rescaling the operators,
$A_i=\sqrt{\Lambda_i/N} B^{\prime}_i$, where $B^{\prime}_i$, which arise by
reshaping the eigenvectors of $ (U^R)^{\dagger}U^R$ according to
(\ref{VSchmidt2}). Reshaped  Kraus operators are eigenvectors of the
hermitian dynamical matrix $D_{\stackrel{\scriptstyle m \mu}{n \nu}}$ of size
$N^2$ which determines \cite{SMR61} the map
\begin{equation}
\rho'\: = \:  \Phi(\rho)
\quad \quad {\rm so} \quad \quad
\rho_{m\mu}\!\!\!\!\!\!'\ \ \: = \:
 D_{\stackrel{\scriptstyle m n}{\mu \nu}}
  \, \rho_{n \nu} \ .
 \label{dynmatr1}
\end{equation}
Taking into account an appropriate normalization we obtain the dynamical
matrix corresponding to the unistochastic map
\begin{equation}
 D_{\Phi_U}= \frac{1}{N}  (U^R)^{\dagger}\,  U^R
\label{dunisto1}
\end{equation}
Hence the rescaled Schmidt coefficients $\Lambda_i/N$ of any unitary matrix
$U$ treated as an element of the composite space of matrices of size $N^2$
provide the spectrum of the dynamical matrix $D$ representing the action of
the unistochastic map  ${\Psi_U}$.

The matrix $D$ is normalized according to Tr$D=\sum_{i=1}^{N^2} \frac{1}{N}
\Lambda_i =N$. Therefore the  {\sl entanglement entropy} (\ref{shannon})
characterizing the non-local properties of $U$ is equal to the entropy of an
operation  $\Phi_U$ \cite{ZB04}. For instance,  the entropy vanishes for any
local $U$ which induces an unitary  ${\Phi_U}$ and  is equal to $2 \ln N$ for
the Fourier matrix (\ref{Four}), corresponding the the maximally depolarizing
channel, ${\Psi_F}(\rho)={\mathbbm 1}/N$.

The environmental representation  (\ref{unistoch}) may be generalized by
allowing a larger size of the environment. Their physical motivation is
simple: not knowing anything about the environment (apart from its
dimensionality), one assumes that it is initially in the maximally mixed
state. In particular, one can define generalized, {\it $K$--unistochastic
maps} determined by a unitary matrix $U(N^{K+1})$, in which the environment
of size $N^K$ is initially in the state ${\mathbbm 1}_{N^K}/N^K$. By
definition, a $1$--unistochastic map is unistochastic.

A map $\Phi$ is called {\sl bistochastic} if it preserves the trace and keeps
the maximally mixed state invariant, $\Psi({\mathbbm 1}/N)={\mathbbm 1}/N$.
The definition (\ref{unistoch}) implies that both these conditions are
satisfied, so any unistochastic map is bistochastic. In the following
sections of this work we demonstrate that the converse is not true.

In the simplest case of one qubit maps any bistochastic map is called a {\sl
Pauli channel}, since it can be brought by means of unitary rotations into
the  form \cite{RSW02}
\begin{equation}
\rho \: \rightarrow \: \rho' \: = \:
 \sum_{i=0}^3
\lambda_i \: \sigma_i \rho\, \sigma_i \  \ \
{\rm \quad with \quad}
\sum_{i=0}^3 \,  \lambda_i =1 \ .
\label{bistPauli}
\end{equation}
Here $\sigma_i$ denote Pauli matrices while $\sigma_0={\mathbbm 1}$. The
Pauli matrices satisfy $\sigma_j=-i\exp(i\pi\sigma_j/2)$, so the extreme
points of the set of the bistochastic maps represent rotations of the Bloch
ball around the corresponding axis by the angle $\pi$.

Describing density matrices through their Bloch vectors, $\rho=\frac{1}{2}
{\mathbbm 1}+ {\vec \tau} \cdot {\vec \sigma}$, we can write any bistochastic
map $\rho' = \Phi \rho$ in the form
\begin{equation}
\vec{\tau}\,'\: =\:
t\vec{\tau}  \: = \:  O_1\eta\,  O^{\rm
T}_2\vec{\tau}  \ .
\label{matrTtau}
\end{equation}

\noindent
Here $t$ denotes a real matrix of size $3$ which we
bring to a diagonal form by orthogonal transformations $O_1$ and $O_2$.
As we permit only unitary rotations of the qubit represented by
orthogonal matrices from $SO(3)$, (reflections are not allowed),
some elements of the diagonal matrix $\eta $ may be negative.
The elements of the diagonal matrix $\eta$ are
called the {\it damping vector} $\vec{\eta} =
(\eta_1, \eta_2, \eta_3)$,
because the transformation (\ref{matrTtau})
takes the Bloch ball to an ellipsoid with three axis
given by $\vec\eta$.

Writing down the superoperator $\Phi$ of
a Pauli channel and
reshuffling this matrix according to (\ref{reshuff})
we obtain the dynamical matrix  $D$.
For any bistochastic maps the
dynamical matrix $D$ splits into two blocks and its eigenvalues are
\begin{eqnarray}
d_{1,4}=\frac{1}{2} \Lambda_{1,4}=\frac{1}{2}[1+\eta_z\pm(\eta_x+\eta_y)]\ , \nonumber \\
d_{2,3}=\frac{1}{2} \Lambda_{2,3}=\frac{1}{2}[1-\eta_z\pm(\eta_x-\eta_y)] \ .
\label{etaddd}
\end{eqnarray}
Due to Choi theorem  the map $\Phi$  is completely positive if
the dynamical matrix $D$ is positive definite.
This is the case if all eigenvalues $d_i=\Lambda_i/2$ are not negative,
which is true for the damping vector $\vec{\eta}$
 satisfying the Fujiwara--Algoet conditions   \cite{FA99}
\begin{equation}
(1\pm \eta_3)^2 \geq (\eta_1\pm \eta_2)^2  \ .
\label{postautau}
\end{equation}
These four inequalities assure that
the corresponding positive map $\Phi_{\vec \eta}$ is CP.
They define a regular tetrahedron whose extreme
 points are $\vec{\eta} = (1,1,1)$, $(1,-1,-1)$,
$(-1,1,-1)$, $(-1,-1,1)$. The first point represents the
identity operation while the three others correspond
to unitary rotations by one of three Pauli matrices.
As shown in section \ref{sec:schmidt-four}
there are  bistochastic maps given by
$\vec{\eta}$ inside the tetrahedron,
for which the representation  (\ref{unistoch}) does not exist,
so they are not unistochastic.

%%====================================================================
\section{Two qubit unitary gates}
\label{sec:canonical-form-two}
%%====================================================================

Treating unitary matrices as quantum gates we need not to
care about an overall phase, since physical states differing
by such a phase are identified. Discussing gates acting on
a bipartite system we may thus fix this phase
and restrict our attention to
matrices pertaining to the special unitary group $SU(N^2)$.

In this section we are going to
analyze the simplest case of two-qubit unitary gates. Hence we set $N=2$
and study matrices of the group $SU(4)$. A choice of two subspaces
distinguishes a  subgroup $SU(2) \otimes SU(2)$. From the physical point
of view it corresponds to selection of two distinct subsystems.

\subsection{Canonical form}

Consider unitary matrices $U$ and $V$ of size $N^2\times N^2$ which act in
the composite Hilbert space ${\cal H}_N \otimes {\cal H}_N$. Two such
matrices are {\sl locally equivalent}, written $U \sim_{\rm loc} V$, if there
exist local operations, $W_A \otimes W_B$ and $W_C \otimes W_D$, such that
$V= (W_A \otimes W_B) U (W_C \otimes W_D)$.

It is known, \cite{KBG01,KC01,Ma02}, that any unitary matrix $U$ of size $4$
is locally equivalent to some matrix of the following {\sl canonical form},
\begin{equation}
  V=\exp\Bigl( i \sum_{k=1}^3 \alpha_k \sigma_k \otimes \sigma_k
  \Bigr):=\exp\Bigl( i H_{int}\Bigr)
  \label{CanU}
\end{equation}
where $\sigma_k$ stand for the Pauli matrices, and $\alpha_k$, are real for
$k=1,2,3$. Here $H_{int}$ denotes a Hermitian operator of order four, called
the {\sl interaction Hamiltonian} of the two-qubit gate.

The two--qubit gate $V$ is periodic in each parameter $\alpha_k$ with period
$\pi/2$. Thus the space of parameters characterizing the gate has a structure
of a $3$-torus $T^3$. However, the coefficients $\alpha_k$ are not determined
unambiguously. Due to the symmetries of the problem out of a cube spanned by
three components of $\vec \alpha$ one can distinguish a smaller subset,
called {\sl Weyl chamber}, such that the correspondence between its points
and  the local orbits is one--to--one \cite{KC01}. For instance,  it is
possible to bring locally the Hamiltonian $H_{int}$ into a form in which
\cite{ZVWS02}
\begin{eqnarray}
%  \frac {\pi}{4} \ge \alpha_1 \ge \alpha_2 \ge \left| \alpha_3\right|  \geq 0.
 \frac{\pi}{4} \ge \alpha_1 \ge \alpha_2 \ge \alpha_3  \geq 0
{\rm \quad or \quad} \\
\frac {\pi}{2} \ge \alpha_1 > \frac{\pi}{4},\  \
\frac{\pi}{2}-\alpha_1 \ge \alpha_2 \ge \alpha_3  \geq 0 .
  \label{canUalpha}
\end{eqnarray}
These restrictions imply that the Weyl chamber has a structure of a
tetrahedron, which forms a $1/24$-th part of the cube -- see Fig 1a.

The three-component vector $\vec\alpha$, called {\sl interaction content} of
the gate, characterize the purely non--local interaction Hamiltonian
$H_{int}$. By construction two unitary gates are locally equivalent if and
only if they are characterized by the same interaction content.

\begin{figure}[htbp]
 % \psfragscanon
 \centering \includegraphics[width=0.48 \textwidth]{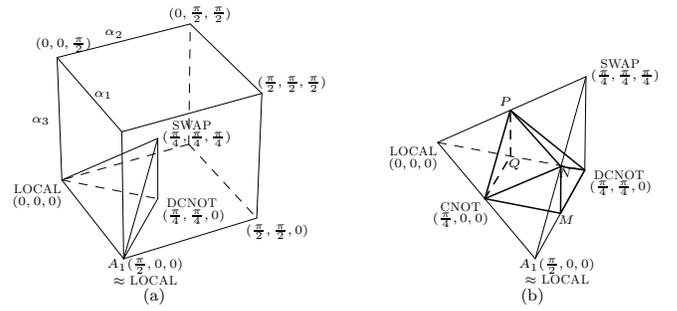}
  \caption{
  a) Any orbit of locally equivalent two--qubit unitary gates
     intersects the Weyl chamber forming a tetrahedron
    in the cube of vectors $\vec \alpha$ representing the
    interaction content. b) The set of perfect entanglers
    forms a polytope inside the Weyl chamber
    with corners at points $P,Q,N,M$,  CNOT and DCNOT  \cite{ZVWS02}.
    Note that point $P$ represents  $\sqrt{\rm SWAP}$.}

  \label{fig:alpha}
\end{figure}

For completeness we present a direct algorithm of finding the information
content  $\vec \alpha$ of any two--qubit unitary gate $U$ based on
\cite{KC01,HVC02}. It relays on the fact from the group theory, $SU(2) \times
SU(2) \sim SO(4)$, which implies that any local operation from $SU(2) \times
SU(2)$ forms in the so--called \emph{magic basis} an orthogonal matrix $O \in
SO(4)$ \cite{HW97}. Therefore any unitary $U$ written in this basis can be
brought to the canonical form  (\ref{CanU}) by an orthogonal rotation,
 \begin{equation}
  \label{OYOT}
  V= O^T U O \ .
  \end{equation}

\noindent
The algorithm works for any $U\in U(4)$ and consist of five steps:

\noindent i) Find $U'=Ue^{-i\chi/4}$ with $\chi$ equal to the phase of the
det$U$, such that $U'\in SU(4)$;

\noindent ii) Write it down in the magic basis, i.e.\ find
$W=MU'M^{\dagger}$, where {\small $M=\frac{1}{\sqrt{2}} \left(
\begin{array}{cccc}
         0  & -i & -i & 0 \\
         1  &  0 &  0 & 1 \\
         -i  & 0 &  0 & i \\
         0  & 1 & -1 & 0
       \end{array} \right)$,
} contains (row--wise) four Bell states forming the  magic basis
$\{-i|\psi^+\rangle,|\phi^+\rangle,-i|\phi^-\rangle,|\psi^-\rangle\}$;

\noindent iii) Compute $WW^T$ and find its spectrum, which we will write in
the form
$\{e^{-2i\delta_1},e^{-2i\delta_2},e^{-2i\delta_3},e^{-2i\delta_4}\}$;

\noindent
iv) Find vector $\vec \delta$  by dividing eigenphases of $WW^T$ by minus two.
 Pay attention  to the total phase: By construction $\sum_i\delta_i$ should
 be equal to zero, so if this is not the case replace $\delta_{max}$ by
 $\delta_{max}-\pi$ or $\delta_{min}$ by $\delta_{min}+\pi$, which
 corresponds to another choice of the signs in $\sqrt{e^{2i\delta}}$.

\noindent
v) Change variables to obtain the information content $\vec \alpha$,
\begin{equation}
\left\{
\begin{array}{ccc}
%\begin{eqnarray}
\alpha_1&=&(\delta_1+\delta_2-\delta_3-\delta_4)/4=(\delta_1+\delta_2)/2,
%\nonumber
\\
\alpha_2&=&(\delta_1-\delta_2+\delta_3-\delta_4)/4=(\delta_1+\delta_3)/2,  \\
\alpha_3&=&(-\delta_1+\delta_2+\delta_3-\delta_4)/4=(\delta_2+\delta_3)/2.
%\nonumber
\end{array} \right.
\label{deltaalpha}
\end{equation}
\medskip

Alternatively,  the vector $\delta$ can be defined as the spectrum of the
Hamiltonian $H_{int}$ entering the canonical form (\ref{CanU}). The
Hamiltonian $H_{int}$ is traceless and diagonal in the magic basis. Its four
eigenvalues $\delta_i$ depend on the information content $\vec \alpha$ in a
linear way. An inverse of (\ref{deltaalpha}) gives
  \begin{equation}
\left\{
    \begin{array}{ccc}
      \delta_1 & = & \alpha_1+\alpha_2-\alpha_3, \\
      \delta_2 & = & \alpha_1-\alpha_2+\alpha_3, \\
      \delta_3 & = & -\alpha_1+\alpha_2+\alpha_3, \\
      \delta_4 & = & -\alpha_1-\alpha_2-\alpha_3, \\
    \end{array}  \right.
    \label{eq:alphadelta}
  \end{equation}
so one may easily switch between both representations.

Note that the existence of  the canonical form (\ref{CanU}) for $N=2$ is due
to the fact that the group $SU(2)$ is homomorphic to $SO(3)$. However, for
higher dimensions $SU(N)$ is homomorphic to a  measure zero,
$(N^2-1)$--dimensional proper subset of the $(N^2-1)(N^2-2)/2$ dimensional
group $SO(N^2-1)$, and  no direct analogue of such a canonical form is known.

\subsection{Schmidt coefficients}
\label{sec:schmidt-four}

By definition (\ref{VSchmidt}) the Schmidt vector ${\vec \lambda}={\vec
\Lambda}/4$ is invariant with respect to local unitary operations. To find
its a relation with the information content $\vec \alpha$ of an arbitrary
unitary matrix of size $N=4$ it suffices to take $U$ in its canonical form
(\ref{CanU}), and find singular values of the reshuffled matrix $U^R$. Their
squares appear in the spectral decomposition of a positive, Hermitian matrix
%$U_{can}^R \left(U_{can}^R \right)^{\dagger}$,

  \begin{equation}
   U_{can}^R \left(U_{can}^R \right)^{\dagger} =
\Lambda_1 \mathbb{I} \otimes \mathbb{I} + \Lambda_2 \sigma_1 \otimes \sigma_1
    + \Lambda_3 \sigma_2 \otimes \sigma_2 + \Lambda_4 \sigma_3 \otimes \sigma_3
\label{uureigen}
  \end{equation}

Simple algebra gives the connection between local invariants:
  \begin{equation}
\!\!\!
\left\{
    \begin{array}{c}
\!\!
  \Lambda_1=     \left(  1+ \cos 2\alpha_2 \cos 2\alpha_3 +\cos  2\alpha_1 \cos 2\alpha_3 +
\cos 2\alpha_1 \cos 2\alpha_2 \right) \ \ \ \\
      \Lambda_2=  \left(  1+ \cos 2\alpha_2 \cos 2\alpha_3 - \cos  2\alpha_1 \cos 2\alpha_3 -
\cos 2\alpha_1 \cos 2\alpha_2 \right) \ \ \ \\
      \Lambda_3=  \left(  1- \cos 2\alpha_2 \cos 2\alpha_3 + \cos  2\alpha_1 \cos 2\alpha_3 -
\cos 2\alpha_1 \cos 2\alpha_2 \right) \  \ \ \\
      \Lambda_4=  \left(  1- \cos 2\alpha_2 \cos 2\alpha_3 -
   \cos  2\alpha_1 \cos 2\alpha_3 + \cos 2\alpha_1 \cos 2\alpha_2 \right) \ ,
    \end{array} \right.
  \label{lambdaccc}
  \end{equation}
Introducing  new variables
  \begin{equation}
\left\{
    \begin{array}{c}
      \eta_1:=\cos 2\alpha_2 \cos 2\alpha_3 \\
      \eta_2:=\cos  2\alpha_1 \cos 2\alpha_3 \\
      \eta_3:=\cos 2\alpha_1 \cos 2\alpha_2
    \end{array}  \right.
\label{eta3}
  \end{equation}
we realize that relations (\ref{lambdaccc}) are equivalent to (\ref{etaddd}).
Thus the coordinates $\vec \eta$ used above represent the damping vector
which enters Eq. (\ref{matrTtau}) and characterizes the corresponding
unistochastic map $\Phi_U$. Note that the vector $\vec \Lambda$ of Schmidt
coefficients of $U$, determines the eigenvalues $d_i=\Lambda_i/N$ of the
dynamical matrix $D_U$ associated with  $\Phi_U$, while
$\lambda_i=\Lambda_i/N^2$ describe the weights in the Pauli channel
(\ref{bistPauli}).

To see which damping vectors $\vec \eta$ correspond to unistochastic maps we
have to describe the image of the Weyl chamber  (\ref{canUalpha}) with
respect to transformation (\ref{eta3}). We may invert this relation
\begin{equation}
\left\{
    \begin{array}{c}
      \alpha_1 = \frac{1}{2}{\rm arccos} \frac{\eta_2\eta_3}{\eta_3} \ \ \\
 \alpha_2 = \frac{1}{2} {\rm arccos} \frac{\eta_1\eta_3}{\eta_2} \ \ \\
 \alpha_3 =\frac{1}{2} {\rm arccos} \frac{\eta_1\eta_2}{\eta_3}\ ,
  \end{array}  \right.
 \label{alphaeta}
  \end{equation}
if the absolute value of arguments of arc cosine are smaller then unity. This
leads to the following restrictions for the set $\cal U$ of damping vectors
corresponding to unistochastic maps
\begin{equation}
 \left\{
  \begin{array}[c]{c}
    \eta_1\eta_2 \leq \eta_3 \ \\
    \eta_2\eta_3 \leq \eta_1 \ \\
    \eta_3\eta_1 \leq \eta_2.
  \end{array} \right.
\label{etaeta}
\end{equation}

A bistochastic one-qubit map $\Phi_{\vec \eta}$ is unistochastic, if the
vector $\vec \eta$ satisfies the set of three conditions (\ref{etaeta}), so
relation (\ref{alphaeta}) gives the interaction content $\vec \alpha$ and the
explicit form (\ref{CanU}) of the unitary matrix defining the map.

\begin{figure}[htbp]
  \centering
  \includegraphics[width=0.42\textwidth{}]{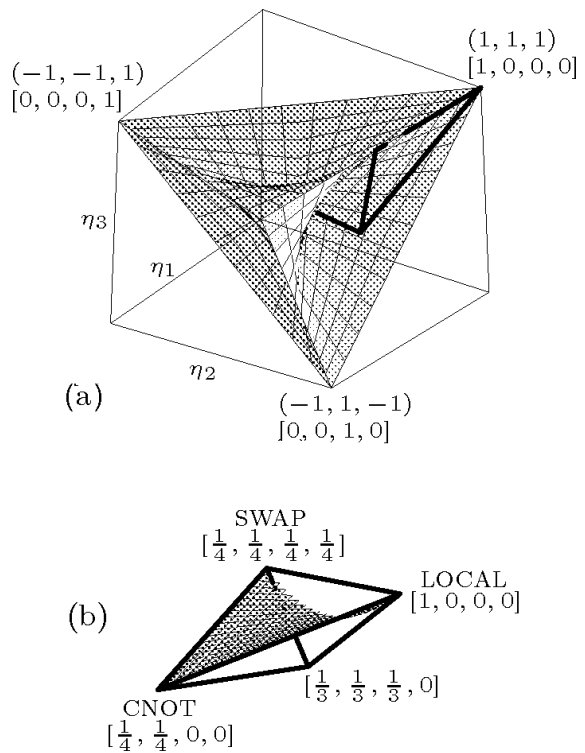}%{unisteta.eps}
  \label{fig:33}
  \caption{a) The set ${\cal U}_2$ of one--qubit unistochastic maps
    forms a proper subset of the tetrahedron of bistochastic maps.
   b) The part of the Weyl chamber of non-equivalent
      unistochastic maps.}
\end{figure}

The set ${\cal U}_2$ of one--qubit unistochastic maps forms a proper subset
of the tetrahedron of bistochastic maps  bounded by three parabolic
hyperboloids $\eta_1=\eta_2\eta_3$, $\eta_2=\eta_1\eta_3$, and
$\eta_3=\eta_1\eta_2$. Note that the set ${\cal U}_2$ is not convex: it
contains all four corners and six edges of the tetrahedron, but its center,
the point $\vec\eta=0$, belongs to the boundary of ${\cal U}_2$ - see Fig 2.
Thus we may model ${\cal U}_2$ by pressing symmetrically four faces of a
tetrahedron with rigid edges in such a way that they all touch in the center
- see Figs. 2 and 3.
%\ref{fig:33} and  \ref{fig:34}.

\begin{figure}[htbp]
  \centering
  \includegraphics[width=0.35\textwidth{}]{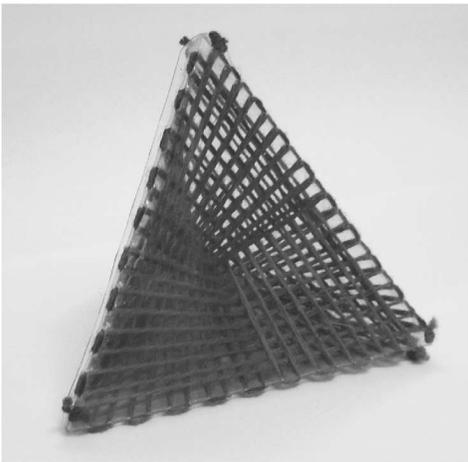}
  \label{fig:34} \caption{A model of the set ${\cal U}_2$ of one--qubit
  unistochastic maps which forms a non--convex subset of the tetrahedron of
  bistochastic maps. The model relays on the fact that the hyperboloids
  (\ref{etaeta}) can be ruled by straight lines.}
\end{figure}

Let us now try to invert relations (\ref{eta3}) and (\ref{alphaeta}) to
obtain the vector $\vec \alpha$ as function of the components of the Schmidt
vector,
\begin{equation}
\left\{
 \begin{array}[c]{c}
\alpha_1= \frac{1}{2} {\rm arccos} \sqrt{ w_1}
{\rm \quad where \quad}
w_1 =
 \frac{(\Lambda_{1}+\Lambda_{4}-2)
(\Lambda_{1}+\Lambda_{3} -2)}
{2(\Lambda_{1}+\Lambda_{2}-2)} \ ,  \\
\alpha_2= \frac{1}{2} {\rm arccos} \sqrt{w_2}
{\rm \quad where \quad}
w_2 =
\frac{(\Lambda_{1}+\Lambda_{4}-2)(\Lambda_{1}+\Lambda_{2}-2)}
{2(\Lambda_{1}+\Lambda_{3}-2)} ,  \\
\alpha_3= \frac{1}{2} {\rm arccos} \sqrt{w_3}
{\rm \quad where \quad}
w_3=
\frac{(\Lambda_{1}+\Lambda_{2}-2)(\Lambda_{1}+\Lambda_{3}
-2)}{2(\Lambda_{1}+\Lambda_{4}-2)} \
 \end{array} \right. .
\label{lambdalpha}
\end{equation}

The choice of the sign of the square root determines the sign of $\alpha_3$.
Constraints for unistochasticity (\ref{etaeta}) imply that $|w_i|\le 1$ for
$i=1,2,3$ so the function arccos is well--defined. Alternatively, these
inequalities  provide constraints $\vec \lambda$ has to satisfy, to represent
a Schmidt vector of a $N=4$ unitary matrix. The choice (\ref{canUalpha})  of
the domain containing $\vec \alpha$ implies that the elements of the Schmidt
vector are ordered non--increasingly, $\Lambda_1\ge \Lambda_2\ge\Lambda_3\ge
\Lambda_4$.

A local unitary operation (${\vec \alpha}=0$) is of rank one, while a generic
two--qubit unitary gate has Schmidt rank $4$. As noted by D{\"ur} and Cirac
\cite{DC02} and Nielsen and al. \cite{NDDGMOBHH02}, there are no unitary
matrices of Schmidt rank $3$.
%In geometric terms, the faces of the Schmidt simplex
%do not belong   to the set of
Indeed, by setting $\Lambda_4$ to zero, equations (\ref{lambdalpha}) impose
$\Lambda_3=0$ as well. Interestingly, for two--qutrit composite systems there
exist unitary gates of size $9$ with Schmidt rank equal to $r$ for each
$r=1,\dots,9$ \cite{Ty03}.

Observe that expressions (\ref{lambdalpha}) are ill defined if all components
are equal, ${\vec \Lambda}_*=(1,1,1,1)$. Hence one may expect that this
degenerated Schmidt vector of the maximal entropy, $S=2\ln 2$, corresponds to
different, non-locally equivalent gates. This fact, discussed in \cite{DC02},
may be demonstrated with explicit examples. Let us define the following
two-qubit unitary gates
\begin{eqnarray}
  U_{\rm CNOT}=
  \left[ \begin{array}
      [c]{cccc}%
      1 & 0 & 0 & 0\\
      0 & 1 & 0 & 0\\
      0 & 0 & 0 & 1\\
      0 & 0 & 1 & 0
    \end{array}
  \right],
  \quad
  U_{\rm DCNOT}=
  \left[ \begin{array}
      [c]{cccc}%
      1 & 0 & 0 & 0\\
      0 & 0 & 0 & 1\\
      0 & 1 & 0 & 0\\
      0 & 0 & 1 & 0
    \end{array}
  \right],
\nonumber \\
  U_{\rm SWAP}=
  \left[ \begin{array}
      [c]{cccc}%
      1 & 0 & 0 & 0\\
      0 & 0 & 1 & 0\\
      0 & 1 & 0 & 0\\
      0 & 0 & 0 & 1
    \end{array}
  \right] .
\quad \quad \quad \quad \quad
  \label{CNOTs}
\end{eqnarray}

Their names are related to applications in the theory of quantum information
\cite{NC00}. The gate  CNOT  performs the {\sl control not} operation:
$U_{\rm CNOT}|a,b\rangle=|a,a \oplus b\rangle$ where $a$ is the control bit,
$b$ is the target bit and the sum is understood modulo two. Defining a
symmetric CNOT operation with the role of bits reversed, $U_{\rm
CNOT'}|a,b\rangle=|a \oplus b,a\rangle$, one defines the {\sl  double CNOT}
gate by the composition $U_{\rm DCNOT}=U_{\rm CNOT} U_{\rm CNOT'}$. The SWAP
gate may be written as a product of three CNOT gates, $U_{\rm SWAP}=U_{\rm
CNOT'} U_{\rm CNOT} U_{\rm CNOT'}$, while its action on any two qubits,
$U_{\rm SWAP}|a,b\rangle=|b,a\rangle$ explains its name. Characterization of
non-local properties of these exemplary two--qubit gates is provided in Table
1, complementary to the data listed in \cite{BS09}. For comparison we include
also  the so--called $B$--gate which interpolates between  CNOT and DCNOT and
is optimal to simulate an arbitrary two--qubit gate \cite{ZVSW04}. %%

%\bigskip
\begin{widetext}
\hskip -1.1cm
\begin{table}[htbp]
\caption{Nonlocal properties of two-qubit unitary quantum gates;
  $s_{\pm}=2 \pm \sqrt{2}$, $t=1/\sqrt{2}$;
perfect entanglers: No (N), Yes -- inside the set (Y), Yes -- at the boundary (B).
}
  \smallskip
\hskip -0.3cm
{\renewcommand{\arraystretch}{1.47}
\begin{tabular}
%[c]{c||c|c|c|}\hline \hline
[c]{l c c c c c c}\hline \hline
Gates &
\parbox{2.0cm}{\centering information \\ content $\vec \alpha$} &
\parbox{2.3cm}{\centering Hamiltonian \\ eigenvalues $\vec \delta$ } &
\parbox{2.1cm}{\centering Schmidt \\ vector $\vec \Lambda$}  &
\parbox{1.0cm}{\centering Schmidt \\ rank} &
\parbox{1.5cm}{\centering damping  \\ vector $\vec \eta$}  &
\parbox{1.5cm}{\centering perfect  \\ entangler} \\
\hline
local gate & $(0,0,0)$ & $(0,0,0,0)$ & $(4,0,0,0)$ & $1$ & $(1,1,1)$ & N \\
$\sqrt{\rm CNOT}$ & $\frac{\pi}{8}(1,0,0)$ & $\frac{\pi}{8}(1,1,-1,-1)$
%  & $(2\!+\!\sqrt{2},2\!-\!\sqrt{2},0,0)$   & $2$ &
& $(s_+,s_-,0,0)$   & $2$ &
% $(1,\frac{1}{\sqrt{2}}, \frac{1}{\sqrt{2}})$
 $(1,t,t)$  & N  \\
CNOT & $\frac{\pi}{8}(2,0,0)$ & $\frac{\pi}{8}(2,2,-2,-2)$ &
 $(2,2,0,0)$ & $2$  & $(1,0,0)$  & B \\
$B$--gate  & $\frac{\pi}{8}(2,1,0)$ & $\frac{\pi}{8}(3,1,-1,-3)$ &
$\frac{1}{2}(3,3,1,1)$  & $4$ & $\frac{1}{2}(1,0,0)$  & Y \\
DCNOT & $\frac{\pi}{8}(2,2,0)$ & $\frac{\pi}{8}(4,0,0,-4)$
& $(1,1,1,1)$ & $4$ & $(0,0,0)$  & B \\
$\sqrt{\rm SWAP}$ & $\frac{\pi}{8}(1,1,1)$ & $\frac{\pi}{8}(1,1,1,-3)$
& $\frac{1}{2}(5,1,1,1)$ & $4$  & $ \frac{1}{2}(1,1,1)$  & B \\
SWAP & $\frac{\pi}{8}(2,2,2)$ & $\frac{\pi}{8}(2,2,2,-6)$
& $(1,1,1,1)$ & $4$ & $(0,0,0)$  & N \\
Fourier & $\frac{\pi}{8}(2,2,-1)$ & $\frac{\pi}{8}(5,-1,-1,-3)$ &
$(1,1,1,1)$  & $4$ & $(0,0,0)$  & N \\
\hline \hline
\end{tabular}
}
\label{tab:gates}
\end{table}
\end{widetext}

The data presented in the table may be easily obtained for an
arbitrary unitary matrix $U$: the spectrum of $UU^R$ gives the Schmidt
vector $\Lambda$, the phases $UU^T$ (normalized in such a way that
their sum vanishes) provide vector $\vec \delta$, while the
information content $\vec \alpha$ follows from relations (\ref{deltaalpha}).

Note that the matrices $U_{\rm DCNOT}$ and $U_{\rm SWAP}$ are invariant with
respect to reshuffling, so the reshuffled matrix is unitary and all four its
singular values are equal to unity.  Reshuffled Fourier matrix remains
unitary, so it has the same singular values. Hence the Fourier matrix, the
gates DCNOT and SWAP are characterized by the same, maximally mixed Schmidt
vector ${\vec \Lambda}_*=(1,1,1,1)$, but they carry different information
content, and thus are not locally equivalent \cite{DC02}. Making use of
(\ref{lambdalpha}) we infer that any gate with the information content ${\vec
\alpha}=(\pi/4,\pi/4,x)$ with an arbitrary $x$ is characterized by the same
maximally mixed Schmidt vector $\Lambda_*$.

For comparison we have provided the data for matrices representing $\sqrt{\rm
CNOT}$ and $\sqrt{\rm SWAP}$, which may be obtained by replacing the
fragments of (\ref{CNOTs}) containing the NOT gate $U_{\rm NOT}= {\small
 \left[ \begin{array}
      [c]{cc}%
      0  & 1 \\
      1 & 0
    \end{array}
  \right] } $
by
$U_{\sqrt{\rm NOT}}:=
{\small \frac{1}{2}
 \left[ \begin{array}
      [c]{cc}%
      1+i  & i-1 \\
      1-i &  1+i
    \end{array}
  \right] } $.
In these cases taking the square root of a gate corresponds to dividing its
interaction content by two, ${\vec \alpha}(U_{\sqrt{\rm SWAP}})=
\frac{1}{2}{\vec \alpha}(U_{\rm SWAP})$. More generally, among several
possibilities of taking a $k$--th root of an unitary matrix $U$ described by
$\vec \alpha$ one can select an unitary matrix $U^{1/k}$ such that its
interaction content equals ${\vec \alpha}/k$.

Let us mention here that the entangling power of integer roots of the SWAP
gate were investigated in \cite{BS08}. Another possibility to characterize
nonlocal properties of unitary gates by so-called  Frobenius fidelity was
recently investigated in \cite{Sa11}.

%%==================================================================
\section{SCOE(4) - ensemble of two-qubit unitary gates }
\label{sec:scoe4}
%%==================================================================

In this section we study random two-qubit unitary gates
described by unitary matrices of size $N=4$.
An ensemble of unitary matrices generated according to the Haar
measure is called \cite{Me91}
{\sl circular unitary ensemble} (CUE).

We aim to derive the probability distributions of local
invariants of unitary matrices of size $4$.
Since quantum states are defined up to a global phase,
we may restrict our attention to the set of {\sl special}
unitary matrices $SU(4)$ with det$U=1$.
An ensemble containing
special unitary matrices with det$U=1$ induced by the Haar
measure will be called SCUE.

To find the distribution $P(\vec \alpha)$
we use magic basis and the representation given by Eq. (\ref{OYOT}).
The Haar measure on CUE(4) induces
in set of symmetric unitary matrices $Y=U U^{T}$
the measure
\begin{equation}
  d_H Y = d_H O^T Y O \label{eq:28} \ .
\end{equation}
invariant with respect to orthogonal similarity $Y\to O^TYO$.
A set endowed with such invariant volume is called
{\sl  circular orthogonal ensemble} (COE),
so the ensemble of symmetric unitary matrices
with determinant equal to unity will be called SCOE.

It is known \cite{Me91}
that random symmetric unitary matrices of  COE
are characterized by the following
joint distribution of eigenphases $\Theta_i$
\begin{eqnarray}
  P(\Theta_1,\Theta_2,\Theta_3,\Theta_4)
 \ = \
{\cal N}
\delta  \left(\Theta_1+\Theta_2+\Theta_3+\Theta_4\right)
\nonumber \\
\prod_{1\leq m<n\leq
    4}\left|e^{i\Theta_n}-e^{i\Theta_m}\right|\ ,
  \label{scoe1}
\end{eqnarray}
where $\cal N$ is the normalization constant.

Integrating over $\Theta_4$ we obtain
\begin{eqnarray}
\label{scoe5}
  P(\Theta_1,\Theta_2,\Theta_3)={\cal N}^\prime
  \left|{e^{i\left(-{\Theta_1}-{\Theta_2}-{\Theta_3}\right)}}
    -{e^{i{\Theta_3}}}\right| \nonumber \\
\times
\left|{e^{i\left(-{\Theta_1}-{\Theta_2}-{\Theta_3}\right)}}
    -{e^{i{\Theta_2}}}\right|
\left|{e^{i\left(-{\Theta_1}-{\Theta_2}-{\Theta_3}\right)}}-{e^{i{\Theta_1}}}\right|
\nonumber \\
\times
  \left|{e^{i{\Theta_3}}}-{e^{i{\Theta_2}}}\right|
  \left|{e^{i{\Theta_3}}}-{e^{i{\Theta_1}}}\right|
  \left|{e^{i{\Theta_2}}}-{e^{i{\Theta_1}}}\right|.
\end{eqnarray}
The angles $\pi/2\geq\Theta_i\geq\pi/2$, $i:=1,2,3,4$ are
the eigenvalues of $Y=UU^T$,
hence
$\Theta_i=2\delta_i$ \cite{HVC02}. Thus
\begin{eqnarray}\label{abc}
  \alpha_1 &=& \frac{1}{4}\left(\delta_1+\delta_2-\delta_3-\delta_4\right)=
  \frac{1}{4}\left(\Theta_1+\Theta_2\right) \\
  \alpha_2 &=& \frac{1}{4}\left(\delta_1-\delta_2+\delta_3-\delta_4\right)=
  \frac{1}{4}\left(\Theta_1+\Theta_3\right) \\
  \alpha_3 &=& \frac{1}{4}\left(-\delta_1+\delta_2+\delta_3-\delta_4\right)=
  \frac{1}{4}\left(\Theta_2+\Theta_3\right),
\end{eqnarray}
i.e.
%\footnote{Uwaga! zmiana oznaczen $(a,b,c)\to \alpha_k$ }
\begin{eqnarray}
  %%\Theta_1&=&2(a+b-c) \\
  %%\Theta_2&=&2(a-b+c) \\
  %%\Theta_3&=&2(-a+b+c).
  \Theta_1&=&2(\alpha_1+\alpha_2-\alpha_3) \\
  \Theta_2&=&2(\alpha_1-\alpha_2+\alpha_3) \\
  \Theta_3&=&2(-\alpha_1+\alpha_2+\alpha_3).
\end{eqnarray}
 In this way we obtain one of the main results of this paper: the joint
probability distribution for the interaction content vector $\vec \alpha$,
induced by the Haar measure on $SU(4)$
\begin{eqnarray}
  P(\alpha_1,\alpha_2,\alpha_3) =&{\cal N}^{\prime\prime}
  \left|\sin\left(2\left(\alpha_1+\alpha_2\right)\right)\right|
  \left|\sin\left(2\left(\alpha_1+\alpha_3\right)\right)\right|
\nonumber \\
\times&  \! \! \! \! \!
\left|\sin\left(2\left(\alpha_2+\alpha_3\right)\right)\right|
  \left|\sin\left(2\left(\alpha_1-\alpha_2\right)\right)\right|
\nonumber \\
\times&  \! \! \! \! \!
  \left|\sin\left(2\left(\alpha_1-\alpha_3\right)\right)\right|
  \left|\sin\left(2\left(\alpha_2-\alpha_3\right)\right)\right| .
\label{m2}
\end{eqnarray}
The resulting normalization constant ${\cal N}^{\prime\prime}$
can be easily calculated to be $2/\pi$ by integrating over the cube
$-\pi\leq\Theta_i\leq\pi$, i.e.\ $-\pi/2\leq c_i\leq\pi/2$, $i=1,2,3$.

Making use of (\ref{eta3})
one can change variables and get the probability distribution
of the damping vector $\vec \eta$,
\begin{equation}
  P(\eta_1, \eta_2, \eta_3)   =
  \frac{\left| \ecc-\ecb \right| \left| \ecb-\eca \right| \left| \eca-\ecc
\right| }
  {\left| 1 - \eca \right| \left| 1 - \ecc \right| \left| 1 - \ecc \right| }
 \label{peta}
\end{equation}

defined for vectors $\vec \eta$ satisfying constraints (\ref{etaeta}).
This expression shows that the density is concentrated in vicinity of the
boundary of the set ${\cal U}_2$ inscribed
inside the tetrahedron of one--qubit bistochastic maps.
Alternatively, one may easily get the  analytical expression $P(\lambda)$
for the density inside the Schmidt simplex
induced by the Haar measure on $SU(4)$,
but we found it more convenient to
work in the $\vec \eta$ representation and to use (\ref{peta}).

This distribution may be applied
to compute average values of various measures
of nonlocality of random two-qubit unitary gate.
Consider, for instance the purity $r$ (locality)
related to the linear entropy of the Schmidt vector
\begin{equation}
  r \ := \ \lambda_1^2+\lambda_2^2+\lambda_3^2+\lambda_4^2
  =\frac{1}{4} \left( 1+ \eta_1^2 +\eta_2^2 + \eta_3^2 \right) \ .
%  =\frac{1}{4} \left(  1+a_1a_2+a_2a_3+a_3a_1\right)
\label{eq:34}
\end{equation}

The average purity of a unitary matrix distributed according to
the Haar measure on $U(N^2)$ was computed by Zanardi \cite{Za01}
\begin{equation}
\langle r \rangle_{N} =
 \frac{2}{N^2+1} ,
\label{ZanRR}
\end{equation}
so in the case of two--qubit gates
this average reads
$\langle r \rangle_{2} =2/5$.

\begin{figure}[htbp]
  \centering
  \includegraphics[width=0.4\textwidth{}]{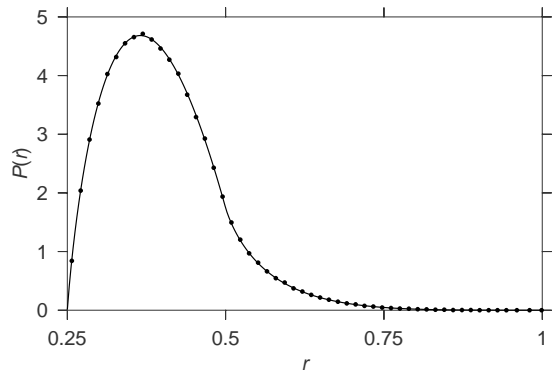}
  \label{fig:pr22}
  \caption{Probability distribution of purity $r$
   of random unitary matrices distributed according to the Haar measure on
   $SU(4)$. Solid line represents numerical integration of distribution
   (\ref{peta}), while black dots show the histogram obtained by a
   Monte-Carlo approach.}
\end{figure}

Making use of the joint distribution (\ref{peta}) we can go a step further
and get the entire probability distribution $P(r)$ as a triple integral. Two
integrals are easy to perform analytically, but the last integral had to be
computed numerically. Results shown as a line in Fig.  \ref{fig:pr22} are
compared with the Monte Carlo calculations in which $10^5$ random unitary
matrices of size $4$ where generated and the distribution of their locality
collected into the histogram denoted by black dots. Results obtained with
both methods agree well and show that a typical random unitary gate has low
purity, so it is strongly non-local. 
There exits a large class of quantum 
gates  with the linear entropy  $r\sim 3/8$, 
close to the value for which the analyzed probability distribution $P(r)$
achieves its maximal value. All these gates represent a 'generic' behavior
in the entire set of two-qubit unitary quantum gates, 
but it is hardly possible to distinguish out of them a 
single gate with some special properties.

%%==================================================================
\section{The subset of perfect entanglers of two-qubit unitary gates }
\label{sec:subs-perf-entangl}
%%==================================================================

Among all two--qubit gates represented by unitary matrices of size $4$, one
may distinguish so--called {\sl perfect entanglers}, which can produce from a
product state a maximally entangled, Bell-like state. A  gate $U$ belongs to
this class if its numerical range (for a normal matrix equal to the convex
hull of the spectrum) includes the eigenvalue $z=0$ \cite{Ma02}.
%has to satisfy, to belong
%to this class was first given by Makhlin \cite{Ma02} and later proved by
%Zhang et al. \cite{ZVWS02}.
In terms of the parameters $\alpha_1,\alpha_2,\alpha_3$
the condition reads \cite{ZVWS02}
\begin{equation}
\label{pe1}
\frac{\pi}{4} - \left |\frac{\pi}{4} - \alpha_1 \right|
\geq   \frac{\pi}{8} - \left |\frac{\pi}{8} - \alpha_2 \right|
\geq \alpha_3 \geq 0,
\end{equation}

In the latter work the authors have shown that the relative volume of the set
of perfect entanglers is equal to $1/2$. However, they used the uniform
measure in the $3$--dimensional space of the vectors representing the
information content (or rather its relevant part, called Weyl chamber), which
does not corresponds the natural Haar measure on the set of unitary matrices.

In this section we redo the calculations using the measure
$P(\alpha_1,\alpha_2,\alpha_3)$ induced by the Haar measure on $U(N)$ and
given by Eq.~(\ref{m2}).

As explained in Section \ref{sec:canonical-form-two} the set of nonequivalent
gates is parametrized by $\alpha_1,\alpha_2,\alpha_3$ restricted by
inequalities (\ref{canUalpha}), i.e.\ belonging to the tetrahedron $T_0$ with
vertices $O=(0,0,0)$, $A=(\pi/2,0,0)$, $U_{\rm DCNOT}=(\pi /4,\pi /4,0)$, and
$U_{\rm SWAP}=(\pi /4,\pi /4,\pi /4)$.
Hence its volume with respect to the measure (\ref{m2}) equals:
\begin{eqnarray}
  V_{w}=&\int\limits_{0}^{\pi/4}d\alpha_1\int\limits_{0}^{\alpha_1}d\alpha_2
  \int\limits_{0}^{\alpha_2}d\alpha_3P\left(\alpha_1,\alpha_2,\alpha_3\right)
 \nonumber \\
\! \! +& \! \! \int\limits_{\pi/4}^{\pi/2} \!
d\alpha_1 \!\! \int\limits_{0}^{\pi/2-\alpha_1} \! \! d\alpha_2
\int\limits_{0}^{\alpha_2}d\alpha_3P\left(\alpha_1,\alpha_2,\alpha_3\right)
=\frac{\cal N}{24} .
\label{Vweyl}
\end{eqnarray}

The set of the perfect entanglers (PE) given by  (\ref{pe1}) is the convex
polyhedron with vertices: $L=(\pi /4,0,0)$, $M=(3\pi /8,\pi /8,0)$,
$U_{\rm DCNOT}$,
$Q=(\pi /8,\pi /8,0)$, $N=(3\pi /8,\pi /8,\pi /8)$, and $P=(\pi /8,\pi /8,\pi
/8)$, and a straightforward calculation of its volume with respect to the
measure (\ref{m2}) gives:
\begin{eqnarray}
  V_{pe} &=&  \frac{\cal N}{9\pi}.
  \label{Vpe}
\end{eqnarray}
Hence the relative volume of the perfect entanglers reads
\begin{equation}
\label{final}
  \frac{V_{pe}}{V_{w}}=\frac{8}{3\pi}\approx 0.85 \ ,
\end{equation}
and gives the probability with which a typical two--qubit gate is
a perfect entangler. This value, much larger than one half,
is consistent with the earlier observations,
that a generic unitary gate is highly non-local.

\begin{figure}[htbp]
  \centering
  \includegraphics[width=0.32\textwidth{}]{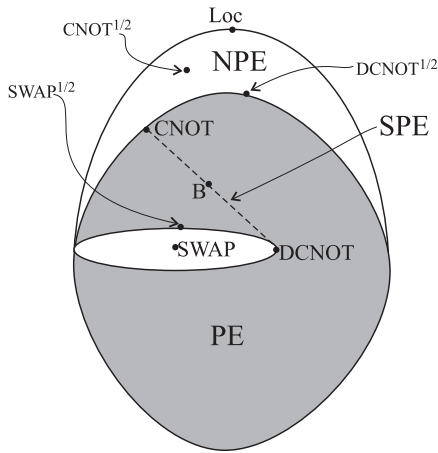}
 \caption{A sketch of the set of two-qubit unitary quantum gates:
 the set of perfect entanglers (PE) occupies approximately
  85\% of the entire volume and contains the set SPE of measure zero.
 The complementary set of non perfect entanglers (NPE) consists
 of two (connected) parts which include local gates (Loc) and SWAP operation, respectively.
}
  \label{fig:sketch}
\end{figure}

In the set of two-qubit gates one distinguishes
also the set of {\sl special perfect entanglers} (SPE),
which can maximally entangle
a set of four product states which form an orthogonal basis \cite{Re04}.
Special perfect entanglers form a one dimensional set of gates interpolating between
CNOT and DCNOT, so the natural measure of this set is equal to zero.
The interpolating gate $B$, introduced in \cite{ZVSW04},
is located inside of the set of PE
and belongs as well to SPE --- see Fig. \ref{fig:sketch},
which shows a sketch of the set of two--qubit unitary gates.

%%==================================================================
\section{ $N \times N$ systems \& Two quNits unitary gates}
\label{sec:n-times-n}
%%==================================================================

Studying non-local properties of a unitary gate $U$
of an arbitrary size $N \times N$
of higher dimensions
 we have to relay on its Schmidt decomposition, since
for $N\ge 3$ no direct analogue of the canonical form (\ref{CanU}) exists.
The Schmidt vector, equal to the squared singular values
of the reshuffled matrix $U^R$,
in several cases may be found analytically.

\subsection{Schmidt vectors for exemplary gates}

 Analyzing, for instance,the general case of the swap operator $U_{\rm SWAP}$
acting now in the space of two quNits, SWAP$|a,b\rangle=|b,a\rangle$, we see
that $U_{\rm SWAP}^R=U_{\rm SWAP}$. Hence the reshuffled matrix is unitary,
so all $N$ of its singular values are equal to unity and the entanglement
entropy is maximal,  $S(U_{\rm SWAP})=2\ln N$. Selecting one element out of
$N$ in each block of the first $N$ rows of $U$ it is straightforward to write
down the other $N!-1$ permutation matrices, which are invariant with respect
to the reshuffling transformation, so their entanglement entropy is maximal.

The same property is also characteristic of the Fourier matrix of order $N^2$
with entries $F_{kl}=\frac{1}{N} \exp(2\pi k l/ N^2)$. Also in this case it
is enough to see that the reshuffled matrix $F^R$ remains unitary, which
implies $S(F)= 2\ln N$. This result was earlier established in
\cite{NDDGMOBHH02,Ty03}. The fact that the entropy of the SWAP gate and of
the Fourier matrix are maximal, does not imply that both gates are locally
equivalent. Up till now the general question which unitary gates of size
$KN>4$ are  locally equivalent remains open.

Among other two--quNits gates let us mention two possible generalizations of
the CNOT (XOR) gate \cite{HH99}
\begin{equation}
U_+|i,j\rangle\equiv |i,i\oplus j\rangle
\quad {\rm and} \quad
U_-|i,j\rangle \equiv |i,i\ominus j\rangle,
\label{GXOR}
\end{equation}
where $i,j=0,1,\cdots,N-1$ and both operations are taken modulo $N$.

Since addition and subtraction of bits modulo two
are equivalent, both gates coincide for
$N=2$ with the standard CNOT gate defined in (\ref{CNOTs}).
The gate $U_-$ may be called a 'controlled rotation',
because the target bit $j$ gets rotated depending on the value of the
control bit $i$ and $(U_-)^d={\mathbbm 1}$.
Inasmuch as the gate $U_+$ is symmetric and forms an involution,
$(U_+)^2={\mathbbm 1}$,
it was called the {\it generalized XOR} gate by
Alber et al. \cite{ADGJ01}, who demonstrated that $U_+$
is capable of performing various tasks of quantum information processing
and proposed a physical realization of this gate
based on non--linear optical elements.

Both gates are permutation matrices with the block diagonal
structure. In both cases all blocks of size $N$ reshaped into vectors
of length $N$ become mutually orthogonal,
so the matrix
$U_{\pm}^R (U_{\pm}^R)^{\dagger}$ is diagonal and consist of exactly
$N$ non-zero elements, each of them equal to $N$. This fact implies that the
entanglement entropy of both gates is equal and reads
\begin{equation}
S(U_{+}) \ = \ S(U_{-}) \ = \ \ln N \ .
\label{SXOR}
\end{equation}

\subsection{Random gates and average values}

The Haar measure on the set of matrices of a composite size $MN$ determines a
natural measure on the set of unitary quantum gates. A gate taken randomly
with respect to this measure will be non--local with probability equal to
one, since the set of local gates of the tensor product structure,
$U_1(N)\otimes U_2(M)$, forms only a zero measure in $U(NM)$. On the other
hand, one may ask to what extend a generic unitary gate is non-local.
Formulating this question more precisely, we will investigate  mean values of
entropies of entanglement averaged with respect to the Haar measure.

The average purity (\ref{ZanRR}) implies that the average linear entropy
$E=1-\sum_{i=1}^{N^2} \lambda_i^2$ reads $\langle E \rangle_{N} =
(N^2-1)/(N^2+1)$ \cite{Za01}. Interestingly, this average is equal to the
mean purity of squared components of a random complex vector of size $N^2$,
distributed according to the natural, unitarily invariant measure on the
space of pure states. However, we will demonstrate that both distributions
are different. In particular, our numerical results show that the average
entropy of entanglement  (\ref{shannon}), behaves like
\begin{equation}
\langle S\rangle_{U}  \sim 2 \ln N -\frac{1}{2} \ ,
\label{SNN}
\end{equation}
while the mean entropy  of a random vector of the size $N^2$ reads
\begin{equation}
\langle S\rangle_{\phi}
=\Psi(N^2+1)-\Psi(2)= \sum_{k=2}^{N^2} \frac{1}{k} \approx 2 \ln N +1-\gamma .
\label{Sjones}
\end{equation}
In this formula, derived in \cite{Jo90}, $\Psi(x)$ denotes the digamma
function while $\gamma\approx 0.5772$ is the Euler constant.

\begin{figure}[htbp]
  \centering
  \includegraphics[width=0.4\textwidth{}]{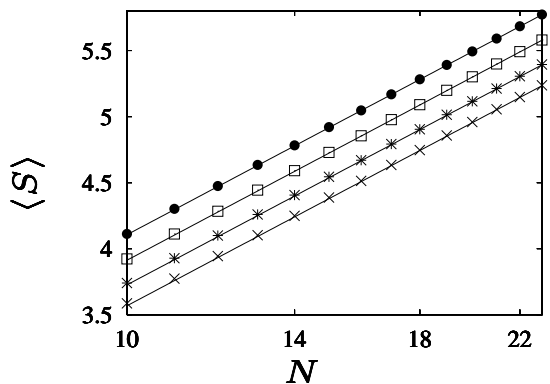}
  \caption{Mean R{\'e}nyi entropies of entanglement averaged over unitary
  matrices acting on symmetric $N \times N$ systems: $(\bullet)$ represents
  the mean Shannon entropy $\langle S \rangle$,  while $(\square)$, $(*)$ and
  $\times$ denote average Renyi entropy  $\langle S_q \rangle $ for $q$ equal to $2,4$ and $8$ respectively.
Dimension $N$ is plotted in the log scale, and  solid lines represent the
asymptotic behavior.}
 \label{fig:meanentr}
 \end{figure}

To characterize statistical distribution of the distribution of Schmidt
vectors of random unitary gates we computed the average moments $\langle
\sum_i \lambda_i^q \rangle$ and the average R{\'e}nyi entropy (\ref{Renyi}).
Our results show that the average entropy for the $N \times N$ unitary gates
behaves as $\langle S_q  \rangle \approx 2\ln N -c_q$
-- see Fig. \ref{fig:meanentr}.  %% 6
Since the dimension $N$ is plotted in the logarithmic scale,
relations (\ref{Sjones}) correspond to straight lines in the graph.
The values of the constants $c_q$ agree with predictions $c_1=1/2$, $c_2=\ln
2\approx 0.69$, $c_4= (\ln 14)/3 \approx 0.88$ derived in \cite{ZS01} for the
average Renyi entropy of mixed quantum states of size $N^2$ obtained by
partial trace of random pure states of an extended system. This fact shows
that a random positive matrix $U^R (U^R)^{\dagger}$ has properties of a
Wishart random matrix: for a large matrix size the unitarity of $U$ hardly
influences statistical property of the corresponding reshuffled matrix $U^R$,
which can be treated as a typical non-Hermitian matrix from the Ginibre
ensemble \cite{Fo10}, as their statistical properties coincide
asymptotically.

A generic quantum pure state  $|\psi\rangle$  of a bipartite $N \times N$
system is typically highly entangled, so the corresponding reduced density
matrix, $\rho={\rm Tr}_N |\psi\rangle \langle \psi|$, is highly mixed and its
von Neumann entropy is close to the maximal value, $\ln N$. This known fact
\cite{ZS01,HLW06} is directly related to properties of random unitary
operation: a generic unitary gate $U$ acting on $N \times N$ system is highly
non-local and the corresponding unistochastic operation $\Psi_U$ is strongly
depolarizing.

Above observations hold also for a general case of $k$-unistochastic channels
or quantum gates acting on $N \times M$ systems. The Haar measure on the
unitary group $U(NM)$ induces a certain measure on the simplex $\Delta_{N^2}$
containing all possible Schmidt vectors  of size $N^2$. The larger size $M$
of the auxiliary subsystem, the larger is the average Shannon entropy of the
Schmidt vector $\Lambda'$, equal to the average entangling power of the
quantum gate. In the limit $M\to \infty$ the average entanglement entropy
tends to the maximal value, $2\ln N$.

\section{Concluding remarks}

In this paper we have analyzed the ensemble of unitary quantum gates
distributed uniformly with respect to the Haar measure on the unitary group.
As the overall phase of the matrix does not influence its action on a quantum
state, we restricted our attention to unitary matrices with determinant equal
to unity and considered matrices from special circular unitary ensemble
(SCUE).

A generic unitary gate $U$, acting on a bipartite $N \times N$ system, was
shown to be strongly non-local. For instance, its entanglement entropy
$S(U)$, (also called Schmidt strength \cite{NDDGMOBHH02}), behaves as $2\ln N
-1/2$, which is also characteristic of random density matrices of size $N^2$,
distributed according to the flat measure \cite{ZS01}.

Any bi-partite quantum gate $U$, represented by a unitary matrix of size
$N^2$, determines a unistochastic operation $\Phi_U$, in which the initial
state of size $N$ is coupled by $U$ with the environment of the same size,
prepared in a maximally mixed state \cite{ZB04}. The entanglement entropy of
a gate $S(U)$, equal to the entropy $S(\Phi_U)$ of the corresponding
unistochastic channel (\ref{unitevol5}), is given by the Shannon entropy of
the eigenvalues of $U^R(U^R)^{\dagger}/N^2$. The reshuffled matrix $U^R$ is
non-Hermitian and for a Haar random unitary $U$, its statistical properties
are shown to coincide with predictions of the Ginibre ensemble. Thus the
spectral density of a normalized, Wishart--like matrix
$U^R(U^R)^{\dagger}/N^2$, is asymptotically described by the
Marchenko--Pastur distribution.

Analyzing in particular the set of two-qubit unitary gates we derived an
explicit formula for the joint probability density (\ref{m2}) for the
interaction content vector $\vec \alpha$ characterizing a quantum gate, and
the distribution (\ref{peta}) for the damping vector $\vec \eta$. Having
these results at hand we were in position to analyze the  subset of
bi-partite quantum gates consisting of perfect entanglers - unitary gates
capable to transform a separable state into a maximally entangled Bell-like
state. Using the known conditions  for a two qubit gate to be a perfect
entangler \cite{Ma02,ZVWS02}, we found that according to the natural Haar
measure on the unitary group  the set of perfect entanglers occupies 
approximately $85\%$ of
the entire volume of the space of unitary matrices of order four. This
observation gives a concrete argument supporting the claim that local gates
are rather exceptional, while a generic unitary gate is strongly non local.

Although any two locally equivalent unitary matrices possess the same set of
the Schmidt coefficients and generate the same unistochastic map, the reverse
is not true. For instance, as shown in Table 1, the following  two-qubit
gates: SWAP, DCNOT and the Fourier matrix are characterized by the uniform
vector $\Lambda$ of the Schmidt coefficients so their entanglement entropy is
equal to $2\ln 2$. Although these unitary matrices are characterized by
different information content $\alpha$, so they are not locally equivalent,
these gates generate the same unistochastic map: the maximally depolarizing
channel.

Any unistochastic operation, determined by a unitary matrix acting on an
extended space, is by construction bistochastic. However, not every
bistochastic operation is unistochastic and can be obtained by a partial
trace over the environment of the same size, initially prepared in the
maximally mixed state. In the case of one qubit maps, there are no
unistochastic maps of rank three, represented by a point belonging to the
face of the tetrahedron of bistochastic maps, spanned by three Pauli matrices
and identity -- see Fig. 3. This is consistent with the known fact that there
is no two--qubit unitary gates of Schmidt rank three \cite{DC02,NDDGMOBHH02}.
For instance the symmetric Pauli channel,
\begin{equation}
\rho \to \rho'= \frac{1}{3} \sum_{i=1}^3 \sigma_i \rho \sigma_i
\label{sympauli}
\end{equation}
 at the center of the face of the tetrahedron of bistochastic maps is located
 as far from the set  ${\cal U}_2$ of one-qubit unistochastic operations as
 possible.

\medskip
It is a pleasure to thank I. Bengtsson and H.--J. Sommers for fruitful
discussions and to N. Johnston for pointing out relevant references.
Support by the grant number N N202-090-239 of
 Polish Ministry of Science and Higher Education
and by SFB/Transregio--12 program financed
by Deutsche Sonderforschungs Gemeinschaft is gratefully acknowledged.

\appendix

%%====================================================================
\section{Schmidt decomposition of a unitary operator}
%\section{Schmidt decomposition and reshuffling}
\label{sec:schm-decomp-resh}
%%====================================================================

\subsection{Matrix algebra: reshaping}

Consider a rectangular matrix $A_{ij}$, $j=1,...,N$ and $i=1,...,M$.
Equivalently, one may put its elements row after row into a vector
$\vec{a}_k$ of size $MN$,
\begin{equation}
  A_{ij}={\vec a}_{(i-1)N+j}.
  \label{matvec}
\end{equation}
For instance, a unitary matrix $U$ of size $M\times M$ will be thus transformed
into a vector $\vec u$ with $M^2$ components.

Let ${\cal H}_N$ denote an $N$ dimensional complex Hilbert space, and ${\cal
H}_{HS}$ the corresponding $N^2$ dimensional Hilbert--Schmidt space of all
linear operators acting on ${\cal H}_N$. It is equipped with a scalar product
$A\cdot B=\langle A|B\rangle:={\rm tr}A^{\dagger}B$, where $A$ and $ B$ are
arbitrary complex matrices of size $N\times N$.

Let $V$ denote an auxiliary unitary matrix of size $N^2$.
Unitarity of $V$ implies that its $N^2$ columns ${\vec{b}}_m=V_{im}$
 (or rows ${\vec{b}}_m=V_{mi}$),
$m,i=1,...N^2$ reshaped into square $N \times N$ matrices $B_m$
as in (\ref{matvec}), form an orthonormal basis in ${\cal H}_{HS}$,
 since $\langle B_m|B_n\rangle:={\rm Tr} B_m^{\dagger}B_n = \delta_{mn}$.
Note that the matrices $B_m$ need not to be unitary.

Consequently, the tensor products $(B_m \otimes B_n)$, $m,n=1,\dots,N^2$, span
an orthonormal basis in the composite Hilbert--Schmidt space ${\cal H}_{HS}
\otimes {\cal H}_{HS} $ of size $N^4$ in which acts
the original $N^2 \times N^2$ unitary matrix $V$.

\subsection{Operator Schmidt decomposition}

Take a given unitary matrix $U$ of size $N^2\times N^2$ we wish to investigate.
It belongs to the composite Hilbert--Schmidt space ${\cal H}_{HS} \otimes {\cal
H}_{HS}$ and will be occasionally denoted as $|U\rangle$. Let us write down its
representation in the basis defined above,
\begin{equation}
|U\rangle =  \sum_{m=1}^{N^2} \sum_{n=1}^{N^2} C_{mn}
 |B_m\rangle\otimes |B_n\rangle,
 \label{Vmatrix}
\end{equation}
where $C_{mn}={\rm tr}( (B_m \otimes B_n)^{\dagger} U)$. The complex matrix $C$
of size $ N^2\times N^2$ needs not to be Hermitian nor normal. The Schmidt
decomposition of $|U\rangle$ reads
\begin{equation}
  |U\rangle = \sum_{k=1}^{N^2} \sqrt{\Lambda_{k}} |B_k^{\prime}\rangle
  \otimes |B_k^{\prime\prime}\rangle,
\label{VSchmidt}
\end{equation}
where $\sqrt{\Lambda_k}$ are the singular values of $C$, (square roots of
eigenvalues of the positive matrix $C^{\dagger}C$) and the basis is transformed
by a local unitary transformation $W_a \otimes W_b$. Thus $ |B^{\prime}\rangle
= W_a |B\rangle$, and $|B^{\prime\prime}\rangle = W_b |B\rangle$ where $W_a$
and $W_b$ are matrices composed of eigenvectors of $C^{\dagger}C$ and
$CC^{\dagger}$, respectively. In a typical case of a non-degenerate spectrum
of $CC^{\dagger}$, the Schmidt decomposition is unique up to two unitary
diagonal matrices, up to which the matrices of eigenvectors $W_a$ and $W_b$ are
determined.

Note that the matrix $C$ depends on the 
initial basis, $\{ B_m \otimes B_n \}_{m,n=1}^N$, in which the analyzed
matrix $U$ is represented, 
while the Schmidt coefficients $\Lambda_k$ are basis independent. 
 Thus, for convenience we may analyze the special
case in which the basis in ${\cal H}_{HS}$ is generated by the identity
matrix, of size $N^2\times N^2$. Then each of the $N^2$ basis matrices $B_n$
of size $N\times N$ has only one non vanishing element which equals unity.
Let's denote $B_{k}=B^{m\mu}=|m\rangle\langle \mu |$, where
{\mbox{$k=N(m-1)+\mu$}}. In this case the matrix of the coefficients $C$ has
a particularly simple form, {\mbox{$C_{\stackrel{\scriptstyle m \mu }{n
\nu}}= {\rm Tr}(B^{m\mu}\otimes B^{n\nu})U= U_{\stackrel{ \scriptstyle
mn}{\mu\nu}}$}}.

\subsection{Matrix algebra: reshuffling}

This particular reordering of a matrix, called {\sl reshuffling} \cite{ZB04},
will be denoted as $U^R:=C$. In general the notion of reshuffling 
is well defined if a matrix $X$ acts on a composite Hilbert space, 
${\cal H}_M \otimes {\cal H}_N$. The symbol $U^R$ has a unique meaning if a
concrete decomposition of the total dimension, $L=MN$,  is specified. Similar
reorderings of matrices were considered by Hill et al. \cite{OH85,YH00} while
investigating CP maps and also  in \cite{Rud02,Rud03,CW03,HHH02,LAS08} to
analyze separability of mixed quantum states.

The Schmidt coefficients of $U$ are equal to the squared singular values
of the reshuffled matrix, $U^R$. Therefore the operator Schmidt decomposition
(\ref{VSchmidt}) of an arbitrary matrix $X$ may be summarized by
\begin{equation}
  \left\{
    \begin{array}{ccl}
      \{ \sqrt{\Lambda_{k}} \}_{k=1}^{N^2} & = &  
% \bigl\{ {\rm SV} (X^R) \bigr\}
 {\rm \ \ singular \ \  values \ \ of \ } X^R
    %  = { \rm  \ square \ \ roots \ \ of \ \ eigenvalues \ \ of  \ }
% (X^R)^{\dagger}X^R
\\
      |B_k^{\prime}\rangle & = &  {\rm \ \ reshaped \ \ eigenvectors \ \ of \ \ }
      (X^R)^{\dagger}X^R   \\
      |B_k^{\prime\prime}\rangle  & = &
      {\rm \ \ reshaped \ \ eigenvectors \ \ of \ \ } X^R(X^R)^{\dagger} \\
    \end{array} \right. \ .
  \label{VSchmidt2}
\end{equation}

Note that the singular values of the reshuffled matrix, ${\rm SV} (X^R)$, are
equal to  square roots  of eigenvalues  of a positive matrix
$(X^R)^{\dagger}X^R$. The initial basis is transformed by a local unitary
transformation $W_a \otimes W_b$, where $W_a$ and $W_b$ are matrices of
eigenvectors of matrices $(X^R)^{\dagger}X^R$ and $X^R(X^R)^{\dagger}$,
respectively. If and only if the rank $K$ of $X^R(X^R)^{\dagger}$ is equal
to one, the operator can be factorized into a product form,
 $X=X_1\otimes X_2$, where $X_1={\rm Tr}_2 X$ and $X_2={\rm Tr}_1 X$.

To get a better feeling of the reshuffling transformation observe that
reshaping each row of the initial matrix $X$ of length $N^2$ according to
(\ref{matvec}) into a submatrix of size $N$ and placing it according to the
lexicographical order block after block produces the reshuffled matrix $X^R$.
Let us illustrate this procedure for the simplest case $N=2$, in which any
row of the matrix $X$ is reshaped into a $2 \times 2$ matrix
\begin{equation}
  C_{kj}=X_{kj}^R :=\left[
    \begin{array}{c|c}
      {\bf {X_{11}\ \ X_{12}}}  & X_{21}{\rm ~ ~ ~ }X_{22} \\
      X_{13} {\rm ~ ~ ~ }X_{14} & {\bf X_{23} \ \ X_{24}} \\
      \hline
      {\bf X_{31}\ \ X_{32}} & X_{41} {\rm ~ ~ ~ }X_{42} \\
      X_{33}{\rm ~ ~ ~ }X_{34} & {\bf X_{43}\ \ X_{44} }
    \end{array}
  \right] .
  \label{reshuf1}
\end{equation}
The operation of reshuffling could be defined in an alternative way.
Instead of reshaping the vectors of $X$ into square matrices of size $N$
%which produces $X^R$, 
one can reshape columns of $X$, which leads to another 
reshuffled matrix $X^{R'}$. 
In the four indices notation introduced above (Roman indices
running from $1$ to $N$ correspond to the first subsystem, Greek indices to
the second one), both operations of reshuffling take the form
\begin{equation}
  X_{\stackrel{ \scriptstyle m \mu }{n \nu}}^{R}:
  =X_{\stackrel{ \scriptstyle m n}{\mu \nu}}
  {\quad \quad  \quad \rm and \quad \quad \quad}
  X_{\stackrel{ \scriptstyle m\mu}{ n \nu}}^{R'}:=
  X_{\stackrel{ \scriptstyle \nu \mu}{n m}}  .
  \label{reshuff}
\end{equation}
However, both reshuffled matrices are equivalent up to a certain permutation of
rows and columns and transposition, so the singular values of $X^{R^{\prime}}$
and $X^R$ are equal. It is easy to see that $(X^{R})^{R}=X$. In general,
$N^{3}$ elements of $X$ do not change their position during the operation of
reshuffling (these typeset {\bf boldface} in (\ref{reshuf1})); the other
$N^{4}-N^{3}$ elements do. The space of complex matrices with the reshuffling
symmetry is thus $2N^4-2(N^4-N^3)=2N^3$ dimensional. Note that if $X$ is
Hermitian the reshuffled matrix $X^R$ needs not to be Hermitian.

\subsection{Entanglement entropy}

The Hilbert-Schmidt norm of any unitary matrix is $||U||=\sqrt{\langle
  U|U\rangle}=\sqrt{N}$. Computing the norm of the right hand side of
(\ref{VSchmidt}) we obtain
\begin{equation}
  \sum_{k=1}^{N^2} \Lambda_k=N^2.
  \label{norm}
\end{equation}
Thus the normalized vector $\vec \lambda$ of the squared singular values,
$\lambda_k:=\Lambda_k/N^2$,
 lives in the ($N^2-1$) dimensional simplex and may
be interpreted as a probability vector. Iff there exists only one non-zero
singular value, $\lambda_1=1$, then the unitary matrix has a product form,
$U=U_a \otimes U_b$. In such a case $U$ is called a {\sl local gate}
and  both operators obtained by partial tracing,
$U_a={\rm Tr}_b U$ and $U_b={\rm Tr}_a U$ are unitary.

In general, the vector of the Schmidt coefficients of an
unitary matrix $U$ acting on a composite $N \times N$ system conveys
information concerning the non-local properties of $U$. To characterize
quantitatively the distribution of $\vec \lambda$ one uses the Shannon entropy,
\begin{equation}
  S(U):= S({\vec \lambda}) = - \sum_{k=1}^{N^2} \lambda_k \ln(\lambda_k)
  \label{shannon}
\end{equation}
called in this context {\sl entanglement entropy} of $U$ \cite{Za01},
(or {\sl Schmidt strength} \cite{NDDGMOBHH02}), and the generalized,
R{\'e}nyi entropies
\begin{equation}
  S_q(U):= S_q({\vec \lambda})  =
-\frac{1}{1-q} \ln \Bigl[ \sum_{k=1}^{N^2} (\lambda_k)^q \Bigr],
  \label{Renyi}
\end{equation}
which tend to $S$ in the limit $q\to 1$. The entropy $S_0$, sometimes called {\sl
Hartley entropy}, is equal to $\ln L$, where $L$ denotes the number of positive
coefficients $\lambda_i$, and is called
 {\sl  Schmidt rank} (or Schmidt number).

The second order Renyi entropy $S_2$ is closely related to the linear entropy
$E(U)=1-\exp(-S_2)$ used by Zanardi in \cite{Za01}. The quantity
$r=\sum_{k=1}^{N^2} (\lambda_k)^2$ is called {\sl purity} while analyzing the
vector of eigenvalues of an arbitrary density matrix $\rho$: the larger
coefficient $r$, the more pure state. The maximal value, $r=1$ is attained if
and only if the state $\rho$ is pure. In the present analysis of the unitary
matrices, we shall stick to this name, although in this context $r$ could be
termed {\sl locality}: the gate $U$ is local if and only if $r=1$.
 Another quantity called
inverse participation ratio is usefull: $R=1/r=\exp(S_2)$ varies from unity
(local gates) to $N^2$ for
%SWAP gates or
 the Fourier unitary matrices of size $N^2$ defined by
\begin{equation}
F_{kl}^{(N^2)}\ := \ \frac{1}{N}\exp\bigl(i2\pi kl/N^2\bigr)\ .
  \label{Four}
\end{equation}
 To demonstrate this fact it is sufficient to notice that the reshuffled matrix
$F^R$ remains unitary, so all its singular values are equal to unity, hence the
Schmidt vector contains $N^2$ equal components and is maximally mixed. Some
examples of unitary two-qubit gates and their Schmidt vector are collected in
Table I.

\subsection{Local equivalence}

By virtue of the Schmidt decomposition if two gates are locally equivalent,
their Schmidt coefficients (and thus the entanglement  entropy)
are equal.
 However, the opposite is not true: there exist unitary
gates with the same set of Schmidt coefficients,
which are not locally equivalent \cite{DC02}.
Hence equality of Schmidt vectors characterizing
two unitary matrices of size $N^2$
is a necessary but not sufficient condition
for their local equivalence.
Sufficient conditions for local equivalence are known
\cite{KBG01,KC01,ZVWS02} only for $N=2$.

\end{document}